\documentclass[12pt]{article}

\usepackage{amsfonts,amssymb,amsmath,amscd}
\usepackage[ps,dvips,matrix,arrow,frame,import,curve,color]{xy}
\newcommand{\ra}{\rightarrow}
\newcommand{\bul}{\bullet}

\newcommand{\CC}{{\mathbb C}}

\newcommand{\bi}{{\bar i}}
\newcommand{\bj}{{\bar j}}

\newcommand{\I}{{\mathcal I}}

\newcommand{\cL}{{\mathcal L}}

\newcommand{\al}{{\alpha}}

\newcommand{\no}{\nonumber}
\newcommand{\id}{{\rm id}}
\renewcommand{\part}{\partial}

\newcommand{\Qb}{Q_{\rm bulk}}
\newcommand{\om}{\omega}

\newcommand{\op}{\oplus}

\newcommand{\bpartial}{{\bar\partial}}

\newcommand{\cI}{{\mathcal I}}
\newcommand{\cJ}{{\mathcal J}}
\newcommand{\cE}{{\mathcal E}}
\newcommand{\cT}{{\mathcal T}}
\newcommand{\cF}{{\mathcal F}}
\newcommand{\bz}{{\bar z}}

\newcommand{\bu}{{\bar u}}
\newcommand{\bw}{{\bar w}}
\renewcommand{\bi}{{\bar i}}
\renewcommand{\bj}{{\bar j}}
\newcommand{\balpha}{{\bar\alpha}}

\newcommand{\bA}{{\bar A}}
\newcommand{\bB}{{\bar B}}
\newcommand{\ot}{\otimes}
\newcommand{\te}{{\tilde e}}
\newcommand{\Ext}{{\rm Ext}}

\title{Open String BRST Cohomology for Generalized Complex Branes}

\author{Anton Kapustin and Yi Li\\{\small \it California Institute of
Technology, Pasadena, CA 91125, U.S.A.}}

\begin{document}

\begin{titlepage}

\maketitle

\begin{abstract}

It has been shown recently that the geometry of D-branes in
general topologically twisted $(2,2)$ sigma-models can be
described in the language of generalized complex structures. On
general grounds such D-branes (called generalized complex (GC)
branes) must form a category. We compute the BRST cohomology of
open strings with both ends on the same GC brane. In mathematical
terms, we determine spaces of endomorphisms in the category of GC
branes. We find that the BRST cohomology can be expressed as the cohomology of a
Lie algebroid canonically associated to any GC brane. In the
special case of B-branes, this leads to an apparently new way to
compute $\Ext$ groups of holomorphic line bundles supported on complex
submanifolds: while the usual method leads to a spectral sequence
converging to the $\Ext$, our approach expresses the $\Ext$ group as
the cohomology of a certain differential acting on the space of
smooth sections of a graded vector bundle on the submanifold. In 
the case of coisotropic A-branes, our computation confirms a proposal
of D.~Orlov and one of the authors (A.K.).

\end{abstract}

\vspace{-6.5in}

\parbox{\linewidth}
{\small\hfill \shortstack{CALT-68-2537}}
\vspace{6.5in}

\end{titlepage}

\section{Introduction and Summary}

More than a decade ago, E. Witten explained how to manufacture 2d topological
field theories from sigma-models whose target is a K\"ahler
manifold~\cite{Witten}. He showed that any sigma-model with a K\"ahler target
space $X$ admits a topologically twisted version called the
A-model; if $X$ is a Calabi-Yau manifold, there is another
topologically twisted theory, the B-model.  Recently it was
realized that A and B topological twists can be applied, under
certain conditions, to more general sigma-models with $(2,2)$
supersymmetry. In these models, the target space is not K\"ahler
in general and the $H$-flux is nonzero. We will call these more
general TFTs the generalized A and B-models. It was shown in
Refs.~\cite{Kap,KLi} that the geometry of these TFTs is
conveniently described in terms of generalized complex geometry~\cite{Hitchin,Gualt}.
Namely, to any $(2,2)$ sigma-model one can associate a pair of
generalized complex structures (see below) $\cI$ and $\cJ$ on the
target space, and the generalized A-model (resp. B-model) depends
only on $\cJ$ (resp. $\cI$), at least if one neglects worldsheet
instantons. Other papers discussing the relation between generalized complex
geometry and supersymmetric sigma-models include Refs.~\cite{LMTZ,CGJ,Bergamin}.

To any topologically twisted sigma-model one can associate a
category of topological D-branes. These are D-branes corresponding
to boundary conditions which are preserved by the the BRST
operator. D-branes associated to the generalized A or B model will
be called generalized complex (GC) branes. The geometry of such
D-branes has been discussed in Refs.~\cite{Kap,Zabz}. We note that
so far only GC branes with abelian gauge fields have been
understood.

By definition, the space of morphisms between a pair of
topological D-branes $\cE$ and $\cE'$ is the BRST cohomology of
the space of open strings, with boundary conditions given by $\cE$
and $\cE'$. From the physical viewpoint, these are simply open
string states in the Ramond sector which have zero energy. Open
string BRST cohomology for topological D-branes in A and B-models
has been intensively studied during the last decade both for
physical and mathematical reasons. In this paper we begin a study
of open string cohomology for GC branes. Our main result is the
description of this cohomology in geometric terms in the case
$\cE=\cE'$. That is, we compute endomorphisms in the category of
GC branes.

Specifically, we show that to any GC brane wrapped on a
submanifold $Y$ of a GC manifold $X$ one can associate a Lie
algebroid $E_Y$ whose cohomology computes the BRST cohomology. It
turns out that this result has interesting implications even for
the well-understood B-branes (i.e. topological D-branes of the
ordinary B-model). Namely, it has been argued that the category of
B-branes is equivalent to the bounded derived category of coherent
sheaves on $X$. Even in the case of a holomorphic line bundle $\cE$
supported on a complex submanifold $Y$, the mathematical procedure
for computing endomorphisms of the corresponding object in $D^b(X)$ is rather complicated
and involves many arbitrary choices. The most explicit way to state the result is to say
that there is a spectral sequence converging to the desired space of endomorphisms whose 
$E_2$ term is given by
\begin{equation}\label{naivespectrum}
\op_{p,q} H^p(\Lambda^q NY).
\end{equation}
The differential $d_2$ can also be described completely explicitly~\cite{KS,Pantev}.
It is the composition of the cohomology class $\beta_Y\in H^1(TY\ot NY^\vee)$ corresponding
to the exact sequence
$$
\begin{CD}
0 @>>>TY @>>> TX\vert_Y @>>> NY @>>> 0
\end{CD}
$$
and the cohomology class $[F]\in H^1(TY^\vee)$ represented by the curvature of the line bundle
$\cE$. The class $[F]$ is known as the Atiyah class of the holomorphic line bundle $\cE$;
it is the obstruction to the existence of a holomorphic connection on $\cE$. The class $\beta_Y$
measures the extent to which $TX\vert_Y$ fails to split holomorphically as $TY\op NY$.
Their composition $[F]\llcorner\beta_Y$ is a class in $H^2(NY^\vee)$.

Back-of-the-envelope estimate of the open-string BRST cohomology
gives $E_2$ as the physical result, but a more careful computation shows
that the whole spectral sequence arises~\cite{KS}. This serves as an important check that the
category of B-branes is indeed equivalent to $D^b(X)$. Our result
shows that one can dispense with the spectral sequence and write
down an explicit graded vector bundle on $Y$ and a differential
$Q_Y$ on its space of sections, such that $Q_Y$-cohomology
computes the space of endomorphisms of the B-brane. Specifically,
the graded bundle is isomorphic to
\begin{equation}\label{grbundle}
\op_{p,q} \Omega^{0,p}\ot \Lambda^q NY^{1,0}
\end{equation}
(the grading being $p+q$),
and the differential $Q_Y$ is mapped by this isomorphism to a deformation of the
Dolbeault differential
$$
\bpartial+\delta(Y,F).
$$
The correction term $\delta(Y,F)$ has bidegree $(2,-1)$ and
depends both on the way $Y$ sits in $X$ and the curvature of the
line bundle on $Y$. The correction term is by itself a
differential, so one can write down a spectral sequence which
converges to $Q_Y$-cohomology and whose $E_1$ term is given by
Eq.~(\ref{naivespectrum}). We show that the differential $d_1$
is equal to $[F]\llcorner \beta_Y$. This confirms that Lie algebroid cohomology
computes $\Ext(\cE,\cE)$.

The isomorphism of our graded bundle on $Y$ with the graded bundle Eq.~(\ref{grbundle})
is not canonical, and as a result the form $\delta(Y,F)$ is not completely
canonical either. However, the construction of the original graded bundle and the
differential $Q_Y$ is completely canonical.

We note that in the case of B-branes of higher rank we do not have
analogous results. The spectral sequence computing endomorphisms
still exists, but we do not know how to get it from a complex of
vector bundles on $Y$. Hopefully, an extension of the computations
in this paper will enable one to find such a complex.\footnote{
We would like to emphasize that the results of Ref.~\cite{KS} apply
only to rank-one bundles.}

\section{Mathematical preliminaries}

Let $X$ be a manifold, and $H$ be a closed 3-form on $X$. The
twisted Dorfman bracket on smooth sections of $TX\op TX^\vee$ is a
bilinear operation $\circ$ defined as follows:
\begin{multline*}
(Z+\xi)\circ (W+\eta)=[Z,W]+\cL_Z\eta-\iota_W d\xi+\iota_Z\iota_W
H,\\ \forall Z,W\in \Gamma(TX),\ \forall \xi,\eta\in
\Gamma(TX^\vee).
\end{multline*}
Its skew-symmetrization is called the twisted Courant bracket. A
subbundle $E$ of $TX\op TX^\vee$ is called integrable if it is
closed w.r. to the twisted Dorfman bracket. Let $q$ be the obvious
symmetric bilinear form on $TX\op TX^\vee$:
$$
q(Z+\xi,W+\eta)=Z(\eta)+W(\xi).
$$
If the subbundle $E$ is isotropic with respect to $q$ and
integrable, then the twisted Dorfman bracket descends to a Lie
bracket on sections of $E$.

A Lie algebroid over $X$ is a triple $(E,[\ ,\ ],a)$, where $E$ is
a real vector bundle over $X$, $[\ ,\ ]$ is a Lie bracket on
smooth sections of $E$, and $a$ is a bundle map $a:E\ra TX$. These
data must satisfy the following requirements:
\begin{itemize}
\item $a([s_1,s_2])=[a(s_1),a(s_2)]$ for any two smooth sections
$s_1,s_2$ of $E$.
\item $[f\cdot s_1,s_2]=f\cdot
[s_1,s_2]-a(s_2)(f) \cdot s_1$ for any two smooth sections
$s_1,s_2$ of $E$ and any $f\in C^\infty(X)$.

\end{itemize}
The simplest Lie algebroid over $X$ is $TX$ itself, with $a=\id$.
All standard constructions using the Lie bracket on $TX$ can be
generalized to an arbitrary Lie algebroid over $X$. For example,
on the sections of the exterior algebra bundle
\begin{equation}\label{extEdual}
\op_p \Lambda^p E^\vee
\end{equation}
there is a degree-1 derivation $d_E$ which squares to zero. Its
definition is a slight generalization of the definition of the
usual exterior derivative:
\begin{multline}
(d_E\alpha)(s_0,\ldots,s_p)=\sum_{i=0}^p (-1)^i
a(s_i)\left(\alpha\left(s_0,\ldots,\widehat{s_i},\ldots,s_p\right)\right)\\
+\sum_{i<j}
(-1)^{i+j}\alpha\left([s_i,s_j],s_0,\ldots,\widehat{s_i},\ldots,\widehat{s_j},\ldots,s_p\right).
\end{multline}
The cohomology of $d_E$ is called the Lie algebroid cohomology.

An integrable isotropic subbundle $E$ of $TX\op TX^\vee$ can be
made into a Lie algebroid by letting $a:E\ra TX$ to be the obvious
projection to $TX$.

A complex Lie algebroid is defined analogously, except $a$ is a
bundle map $a: E\ra TX_\CC$.  If $E$ is an integrable isotropic
subbundle of the complexification of $TX\op TX^\vee$, then it has
an obvious structure of a complex Lie algebroid. From now on we
will drop the adjective ``complex''; since we will be dealing only
with complex Lie algebroids in this paper, this cannot lead to
confusion.

Let $X$ be a manifold and $H$ be a closed 3-form on $X$. A twisted
generalized complex (GC) structure on $(X,H)$ is an endomorphism
$$
\cI: TX\op TX^\vee\ra TX\op TX^\vee,
$$
such that $\cI^2=-1$, $\cI$ preserves $q$, and the eigenbundle of $\cI$ with eigenvalue $-i$ is integrable.
We will denote the latter bundle $E$ in the rest of the note. It is obviously isotropic, so we get a
complex Lie algebroid for every GC structure $\cI$ (we will neglect to say ``twisted'' in the rest of the
note, so ``GC'' will mean ``twisted GC'').

The simplest examples of GC structures (with $H=0$) are given by complex and symplectic structures.
Given a complex structure $I$ on $X$ (regarded as an endomorphism of $TX$), we let, in an obvious
notation,
$$
\cI=\begin{pmatrix} I & 0 \\ 0 & -I^\vee \end{pmatrix}.
$$
Given a symplectic structure $\omega$ on $X$, we let
$$
\cI = \begin{pmatrix}0&-\om^{-1}\\ \om&0\end{pmatrix}.
$$
One can easily check that these tensors define GC structures on $X$~\cite{Hitchin}.

In this paper we will find the following equivalent definition of
a Lie algebroid useful~\cite{Vaintrob}. Given a vector bundle $E$
we can construct a graded supermanifold $E[1]$ by declaring the
linear coordinates on the fibers of $E$ to be fermionic variables
of degree $1$. A Lie algebroid over $X$ is a pair $(E,Q)$, where
$E$ is a vector bundle on $X$ and $Q$ is a degree-1 vector field
on the supermanifold $E[1]$ satisfying
$$
Q^2=0.
$$
To see the relation between the two definitions, let $e_a$ be a
local basis of sections of $E$, let $x_i$ be local coordinates on
$X$, and let $\theta^a$ be fermionic linear coordinates on the
fibers of $E$ dual to $e_a$. The most general vector field on $E[1]$ of degree
$1$ has the form
$$
Q=a^i_\alpha \theta^\alpha
\partial_i+c^\alpha_{\beta\gamma}\theta^\beta\theta^\gamma\frac{\partial}{\partial\theta^\alpha}
$$
for some locally-defined functions $a^i_\alpha,c^\alpha_{\beta\gamma}$. We can use
these functions to define a bundle map $a:E\ra TX$ and a bracket
operation on sections of $E$ by letting
$$
a(e_\alpha)=a^i_\alpha \partial_i,\quad
[e_\beta,e_\gamma]=c^\alpha_{\beta\gamma} e_\alpha.
$$
The condition $Q^2=0$ is equivalent to the requirement that the
triple $(E,[,],a)$ be a Lie algebroid. Note that in this
alternative formulation sections of the graded bundle
Eq.~(\ref{extEdual}) are regarded as functions on $E[1]$, and the
Lie algebroid differential $d_E$ is simply the derivative of a
function along the vector field $Q$.

\section{The Lie algebroid of a GC brane}

A brane of rank one is a submanifold $Y$ together with a
Hermitian line bundle $\cE$ equipped with a unitary connection
$\nabla$. Its curvature $F=-i\nabla^2$ is a real closed 2-form on
$Y$ whose periods are integral multiples of $2\pi$. In what follows,
only the curvature of the connection $\nabla$ will be important;
for this reason we will regard as rank-1 brane as a pair $(Y,F)$.

If $H\neq 0$, then there is an additional constraint of $Y$: the
restriction of $H$ to $Y$ must be exact. That is, while the
B-field on $X$ is not a globally well-defined 2-form, its
restriction to $Y$ is. The set of B-fields on $X$ is acted upon by
1-form gauge transformations:
$$
B\mapsto B+d\lambda,\quad \lambda\in \Omega^1(X).
$$
Under this gauge transformation, the connection on $\cE$
transforms as follows:
$$
\nabla\mapsto \nabla-i\lambda\vert_Y.
$$
The curvature of $\nabla$ is not invariant under these
transformations; the gauge-invariant combination is
$$
\cF=B\vert_Y+F.
$$

The generalized tangent bundle $\cT Y_F$ of a brane $(Y,F)$ will
be defined as the subbundle of
$$
\left(TX\op TX^\vee\right)\vert_Y
$$
defined by the following condition:
$$
Z+\xi\in \cT Y_F\Longleftrightarrow Z\in TY,\ \xi+\iota_Z \cF\in
NY^\vee.
$$

Let $(X,H)$ be a GC manifold with a GC structure $\cI$. A GC brane
on $X$ is defined to be a brane $(Y,F)$ such that its generalized
tangent bundle $\cT Y_F$ is preserved by $\cI$. It was shown in
Ref.~\cite{Kap} that rank-1 topological branes of the
generalized B-model are precisely GC branes.

The definition of a GC brane simplifies somewhat when $\cF=0$,
because its generalized tangent bundle becomes the sum of the
tangent and the conormal bundle of $Y$. We will call a GC brane
with $\cF=0$ a GC-submanifold.

Let $(Y,F)$ be a GC brane in $(X,H,\cI)$. Let $E_Y$ be the $-i$
eigenbundle of the restriction of $\cI$ to $\cT Y_F$. $E_Y$ is a
subbundle of the complefixication of $\cT Y_F$. It turns out there
is a natural Lie algebroid structure on $E_Y$.\footnote{This Lie
algebroid structure was independently found by M. Gualtieri.} The
anchor map is the obvious projection to $TY_\CC$. The Lie bracket
is defined as follows. Given any two sections of $E_Y$, we can
regard them as sections of $E\vert_Y$ (because $E_Y$ is a
subbundle of $E\vert_Y$). Extend them off $Y$, compute the twisted
Dorfman bracket, and restrict back to $Y$. One can easily check
that the result lies in $E_Y$, and does not depend on how we
extend sections off $Y$.

\section{Open-string BRST cohomology for GC branes}

In this section we show that the cohomology of the Lie algebroid
$E_Y$ is isomorphic (classically, i.e. if one neglects instantons)
to the BRST cohomology of the open string space of states, where
both ends of the open string are on the brane $(Y,F)$.

The proof is a combination of two tautological lemmas. The first
one is that open-string BRST cohomology is isomorphic to the
cohomology of a degree-1 vector field $Q_Y$ on a certain graded
supermanifold of the form $L[1]$, where $L$ is some complex vector bundle over
$Y$. Indeed, in the zero-mode approximation (which is sufficient
for computing the BRST cohomology) open-string pre-observables are
functions of both bosonic coordinates on $Y$ and fermionic
coordinates taking values in some vector bundle over $Y$.
Fermionic coordinates can have R-charge $1$ or $-1$. In order to
compute the BRST cohomology, it is sufficient to consider
fermionic coordinates with R-charge $1$, since the BRST-variation
of the ones with R-charge $-1$ contains spatial derivatives of
bosonic coordinates. Let $L$ be the vector bundle over $Y$ where
fermionic coordinates of charge $1$ take values. Then the space of
observables is the space of functions on the graded supermanifold
$L[1]$. The generator of the BRST transformation is a degree-1
vector field on $L[1]$ which squares to zero.

We can be more specific about the bundle $L$. For closed strings,
the fermionic zero modes $(\psi_+,\psi_-)$ take values in the
bundle $TX\op TX$. To make contact with generalized complex
geometry, it is useful to work with their linear combinations
which take values in $TX\op TX^\vee$~\cite{Kap}: $$
\psi=\frac{1}{2}\left(\psi_++\psi_-\right),\quad
\rho=\frac{1}{2}g\left(\psi_+-\psi_-\right),
$$
where $g$ is the Riemannian metric on $X$. Open-string boundary
conditions put a linear constraint on the fermionic zero modes
$(\psi,\rho)$ which requires them to be in the fibers of the
generalized tangent bundle of the brane $(Y,F)$~\cite{Kap}.
Finally, the requirement that the R-charge of the fermions be $1$
is equivalent to the requirement that the fermions be in the
subbundle $E_Y$~\cite{KLi}. Thus $L=E_Y$.

The vector field $Q_Y$ on $L[1]=E_Y[1]$ can be thought of as
follows. In the closed-string case, bosonic zero-modes take values
in the whole $X$, while fermionic zero-modes with R-charge $1$
take values in the bundle $E$. The closed-string BRST operator
$\Qb$ can be thought of as a degree-$1$ vector field on $E[1]$.
Open-string boundary conditions select a submanifold $E_Y[1]$ of
$E[1]$. In these terms, compatibility of the boundary condition
with the BRST symmetry means that $\Qb$ is tangent to $E_Y[1]$.
Therefore $\Qb$ induces a degree-1 vector field $Q_Y$ on functions
of $E_Y[1]$. Obviously, this vector field generates BRST
transformations of open-string pre-observables.

The second tautological statement is that the Lie algebroid
cohomology of $E_Y$ is isomorphic to the cohomology of $Q_Y$
acting on functions on $E_Y[1]$. This is fairly obvious from the
way the Lie bracket on $E_Y$ was defined. Suppose $f$ is a
function on $E_Y[1]$. To compute $Q_Y(f)$, we must extend $f$ to a
function on the ambient supermanifold $E[1]$, apply $\Qb$ and
restrict back to $E_Y[1]$. We can think of the extension as a
two-step procedure. First we extend in the fermionic directions.
This means that if we regard $f$ as a section of $\Lambda^\bul
E_Y^\vee$, we must lift it to a section $\tilde f$ of
$\Lambda^\bul E^\vee\vert_Y$ (the former vector bundle is quotient of the latter).
Second, we extend in the bosonic
directions. This means that we extend the section $\tilde f$ of
$\Lambda^\bul E^\vee\vert_Y$ off $Y$. Then we apply the vector
field $\Qb$, restrict back to $Y$, and project to $\Lambda^\bul
E_Y^\vee.$ One can easily see that these are precisely the
manipulations one has to do to compute the action of the
Lie-algebroid differential for $E_Y$ on a section $f$ of
$\Lambda^\bul E^\vee_Y$.

\section{Examples}

\subsection{GC submanifolds}

It is easy to verify that for GC submanifolds (i.e. for $\cF=0$)
the application of the results of the previous section gives familiar results. 
For a GC structure $\cI$ coming from a
complex structure $I$ on $X$ (and $B=0$), a GC submanifold is
simply a complex submanifold. For such a submanifold,
$E_Y=TY^{0,1}\op \left(NY^\vee \right)^{1,0}$. The Lie bracket is
the obvious one: $TY^{0,1}$ has the standard Lie bracket, the
conormal part is an abelian subalgebra, and $TY^{0,1}$ acts on
sections of $\left(NY^\vee \right)^{1,0}$ via the ordinary
$\bpartial$ operator. Lie algebroid cohomology of $E_Y$ is
therefore isomorphic to
$$
\op_{p,q} H^p\left(\Lambda^q NY^{1,0}\right).
$$

If $\cI$ comes from a symplectic structure on $X$ (and $B=0$),
then a GC submanifold is simply a Lagrangian submanifold, and
$E_Y$ is isomorphic to $TY_\CC$ as a Lie algebroid. Hence Lie
algebroid cohomology is isomorphic to the de Rham cohomology
$H^\bul(Y,\CC)$.

\subsection{Rank-one B-branes}

Now let $B=0$ and let $\cI$ come from a complex structure on $X$.
Let $(Y,F)$ be an arbitrary GC brane. This means that $Y$ is a
complex submanifold of $X$, and the curvature $F$ of the
connection $\nabla$ is of type $(1,1)$ (i.e. the line bundle $\cE$
is holomorphic). Let us compute the Lie algebroid $E_Y$ and show
that it is a deformation of $TY^{0,1}\op \left(NY^\vee
\right)^{1,0}$. Let $(z^\alpha,u^i)$ be local holomorphic
coordinates on $X$ such that $Y$ is locally given by the equations
$z^\alpha=0$. Their complex-conjugates will be denoted
$\bz^\balpha,\bu^\bi$. We want to choose a local basis of sections
for $E_Y$. The most obvious choice is
$$
e_\bi=\frac{\partial}{\partial \bu^\bi} - F_{\bi j} du^j,\quad  e^\alpha=dz^\alpha.
$$
It is easy to see that this is a local trivialization of $E_Y$.
Moreover, it is easy to check that all Lie brackets vanish (it
is important here that $dF=0$). On the other hand, the obvious Lie
algebroid $E_Y^0=TY^{0,1}\op \left(NY^\vee \right)^{1,0}$ has the
following obvious local trivialization:
$$
f_\bi=\frac{\partial}{\partial \bu^\bi} ,\quad f^\alpha=dz^\alpha.
$$
Obviously, all Lie brackets vanish as well. It seems at this stage
that we have proved that the two Lie algebroids are isomorphic.
However, this conclusion is premature, because the transition
functions in the two cases are different. Namely, as one goes from
chart to chart, the covectors $du^j$ mix up with $dz^\alpha$, and so
$e_\bi$ mix with $e^\alpha$; on the other hand, $f_\bi$ does not mix
with $f^\alpha$.

To compare the two Lie algebroids, it is convenient to choose a more complicated local trivialization
for $E_Y$, so that the transition functions are the same as for $E_Y^0$. This will prove that $E_Y$ and
$E_Y^0$ are isomorphic as vector bundles. However, we will see that they are not isomorphic as Lie
algebroids, in general, because some Lie brackets in the new basis will be nonvanishing.

Suppose we have two overlapping charts. The holomorphic coordinates in the other chart will be
denoted $(y^\alpha,w^i)$. We have:
$$
\frac{\partial}{\partial \bu^\bi}=\bA^\bj_\bi \frac{\partial}{\partial \bw^\bj},\quad
du^j=B^j_i dw^i+C^j_\alpha dy^\alpha,
$$
where $B=\left(A^{-1}\right)^t$ is a square matrix whose entries are holomorphic functions of $y$ (it is the
gluing cocycle for $TY^\vee$), $\bA$ is the complex-conjugate of $A$, while $C$ is a rectangular matrix whose entries are holomorphic functions
of $w,z$. On the overlap of the two charts, consider the following holomorphic section of $TY^{1,0}\ot
\left(NY^\vee\right)^{1,0}$:
$$
\gamma=C^j_\alpha \frac{\partial}{\partial u^j}\ot dy^\alpha.
$$
$\gamma$ is a Cech 1-cocycle with values in the coherent sheaf
$TY^{1,0}\ot \left(NY^\vee\right)^{1,0}$ which measures the
failure of $TX\vert_Y$ to split holomorphically as $TY\op NY$. Its class was denoted $\beta_Y$ in
the Introduction.
Using a partition of unity, we can write $\gamma$ as a coboundary
of a smooth 0-cocycle:
$$
\gamma=p^j_\alpha \frac{\partial}{\partial u^j}\ot dz^\alpha -q^j_\alpha\frac{\partial}{\partial w^j}\ot
dy^\alpha,
$$
where the matrices $p^j_\alpha$ and $q^j_\alpha$ are defined on
the first and second chart, respectively, but are not holomorphic,
in general.

Now consider a modified local trivialization: on the first chart
we use
$$
\frac{\partial}{\partial \bu^\bi}-F_{\bi j} \left(du^j-p^j_\alpha dz^\alpha\right),\quad
dz^\alpha,
$$
while on the second chart we use
$$
\frac{\partial}{\partial\bw^\bi}-G_{\bi j}\left(dw^j-q^j_\alpha dy^\alpha\right),\quad dy^\alpha.
$$
Here
$$
G_{\bi j}=\bB_\bi^\bj F_{\bj i} B^i_j
$$
is the matrix representing the 2-form $F$ in the coordinate basis $dw^i,d\bw^\bi$. It is easy to check
that the gluing cocycle between the modified local bases is exactly the same as for
$TY^{0,1}\op\left( NY^\vee\right)^{1,0}$. Thus $E_Y$ is isomorphic to the latter as a vector bundle.

Now let us compute the Lie brackets of the elements of the modified local basis of $E_Y$.
In terms of the old basis, the new one is
$$
\te_\bi=e_\bi+F_{\bi j} p^j_\alpha dz^\alpha,\quad \te^\alpha=e^\alpha=dz^\alpha.
$$
It follows that the Lie brackets on the first chart are
$$
[\te_\bi,\te_\bj]=\left(F_{\bj k}\bpartial_\bi p^k_\alpha-F_{\bi k}\bpartial_\bj p^k_\alpha\right)\te^\alpha,\quad
[\te^\alpha,\te^\beta]=0,\quad [\te_\bi,e^\alpha]=0,
$$
and similarly on the second chart (with $p^k_\alpha$ replaced with
$q^k_\alpha$). We observe that the commutator of $\te_\bi$ and
$\te_\bj$ differs from the commutator of
$f_\bi=\bpartial_\bi$ and $f_\bj$ by a term
$$
\delta(Y,F)(\bpartial_\bi,\bpartial_\bj),
$$
where $\delta(Y,F)$ is a local section of $\Omega^{0,2}(Y)\ot\left(NY^\vee\right)^{1,0}$ given in
the first chart by
$$
\delta(Y,F)=\left(F_{\bj k}\bpartial_\bi p^k_\alpha-F_{\bi k}\bpartial_\bj p^k_\alpha\right)dz^\alpha\ot d\bz^\bi\wedge d\bz^\bj.
$$
This section is actually globally well-defined: this follows from the definition of $p,q$ and
the fact that the 1-cocycle $\gamma$ is holomorphic. To see this more clearly, note that
on the overlap of the two charts we have
$$
\bpartial_\bi p^k_\alpha \frac{\partial}{\partial u^k}\ot dz^\alpha=\bpartial_\bi q^k_\alpha \frac{\partial}{\partial w^k}\ot dy^\alpha.
$$
Here $\bpartial_\bi$ denotes either $\frac{\partial}{\partial \bu^\bi}$ or
$\frac{\partial}{\partial \bw^\bi}$ (they are related by the matrix $\bA$). Thus we have a global
section $d_Y$ of $\Omega^{0,1}(Y)\ot TY^{1,0}\ot \left(NY^\vee\right)^{1,0}$ whose local expression is
$$
d_Y=\frac{\partial p^k_\alpha}{\bpartial \bu^\bi}\, \frac{\partial}{\partial u^k} \ot
dz^\alpha\ot d\bu^\bi.
$$

It is easy to see that $d_Y$ is $\bpartial$-closed and therefore represents
a class $\beta_Y\in H^1(TY\ot NY^\vee)$. This is simply the Dolbeault representative of the cohomology 
class whose Cech representative was denoted $\gamma$.
The form $\delta(Y,F)$ is obtained by taking the wedge product
of $d_Y$ and $F\in\Omega^{0,1}\ot \left(TY^\vee\right)^{1,0}$ and contracting $TY^{1,0}$ with
$\left(TY^\vee\right)^{1,0}$. Since both $F$ and $d_Y$ are $\bpartial$-closed, so is $\delta(Y,F)$.

The Lie algebroid differential for $E_Y$ is now easily computed.
Since $E_Y\simeq TY^{0,1}\op\left( NY^\vee\right)^{1,0}$, it is a
degree-1 differential $Q_Y$ acting on smooth sections of the
graded bundle
$$
\op_{r,s}\Omega^{0,r}(Y)\ot\Lambda^s NY^{1,0},
$$
where the grading is given by $r+s$. One easily sees that if $\zeta$ is a section of this graded bundle,
then
$$
Q_Y(\zeta)=\bpartial \zeta+\delta(Y,F)\llcorner\, \zeta .
$$
Here $\llcorner$ means contraction of $NY^\vee$ and $\Lambda^s
NY$. We conclude that the Lie algebroids $E_Y$ and $E_Y^0$ are not
isomorphic, in general: the former is a deformation of the latter.

Note that the sheaf cohomology class represented by $\delta(Y,F)$
is exactly the product of the class $\beta_Y\in H^1(TY\ot NY^\vee)$
and a class in $H^1(TY^\vee)$ represented by the $(1,1)$ form $F$.
The latter class is the Atiyah class of the line bundle on the
brane $Y$. Thus for $Q_Y$-cohomology we get a spectral sequence
whose first term ($E_1$) is simply the $\bpartial$-cohomology:
$$
\op_{p,q} H^p(\Lambda^q NY),
$$
and the first differential is the product of $\left[d_Y\right]=\beta_Y$
and the Atiyah class. This is precisely the $E_2$ term in the
spectral sequence computing the $\Ext$ groups of the object of
$D^b(X)$ corresponding to our brane $Y$~\cite{KS,Pantev}. 
(The object
is the push-forward of the locally free sheaf $\cE$ on $Y$ to the
ambient manifold $X$). This provides some evidence that the Lie
algebroid cohomology computes the $\Ext$ groups.

\subsection{Coisotropic A-branes}

Since the geometry of coisotropic A-branes is somewhat more
complicated than that of B-branes, we start with a brief review of the
data involved (see Ref.~\cite{KO} for more details). A
coisotropic A-brane is a triple $(Y,\nabla,F)$ such that $Y\subset
X$ is a coisotropic submanifold, and $\nabla$ is a unitary
connection on a line bundle on $Y$ with curvature $F$. By
definition, $\cL Y\equiv\ker(\om|_Y)$ forms an integrable
distribution of constant rank, which is the codimension of $Y$. In
addition, the curvature form $F$, regarded as a bundle map
$F:TY\to TY^\vee$, must annihilate $\cL Y$. So if we denote the quotient bundle $TY/\cL Y$ by $\cF Y$, $F$ descends to a section
of $\wedge^2 \cF Y^\vee$. Finally, the restriction of $\om^{-1}F$ to $\cF Y$ defines a transverse 
almost-complex structure on $Y$ with respect to the foliation $\cL Y$. (This transverse almost-complex structure
is automatically integrable.) It follows from these conditions that the complex
dimension of $\cF Y$ is even. Furthermore, both $F$ and $\om|_{\cF Y}$ are of type $(2,0)+(0,2)$ with respect to the transverse complex structure $J=\om^{-1}F|_{\cF Y}$.

The Lie algebroid associated with the brane $(Y,F)$ is $E_Y=\ker(\I_Y+i)$, where $\I_Y$ is the restriction to $Y$
of the generalized complex structure associated to the symplectic structure $\om$ on $X$:
$$\I = \begin{pmatrix}0&-\om^{-1}\\ \om&0\end{pmatrix}.$$
It is easy to see that topologically $E_Y$ is isomorphic to $\cL_\CC
Y\oplus \cF Y^{1,0}$. However, we will show that $E_Y\simeq \cL
Y_\CC\oplus \cF Y^{1,0}$ also as Lie algebroids. To this end, we will
perform the same kind of calculation as in the B-brane case. Let
us choose a local system of coordinates $(x^a, z^i,\bar{z}^\bi,
y^\mu)$ on $X$ such that the submanifold $Y$ is locally defined by
$x^a=0$, $y^\mu$ parametrize the leaves of the foliation defined
by $\cL Y$, while the $z$'s are holomorphic coordinates in the
transverse directions. Note that the splitting of transverse
coordinates into holomorphic and anti-holomorphic ones is done
with respect to the complex structure $J$ on $\cF Y$. A local
trivialization for $E_Y$ is given by
$$e_i=\frac{\part}{\part z^i}-i\om_{ij}dz^j-i\om_{ia}dx^a, \qquad e_\mu = \frac{\part}{\part y^\mu} -
i\om_{\mu a}dx^a.$$

To find the gluing cocycle in this basis, let us take another
local system of coordinates $(u,w,v)$ which overlaps with the old
one. As equations $x^a=0$ and $u^a=0$ define the same submanifold
$Y$ locally, one must have $u=u(x)$ on the overlap. In addition,
from $\om(\cL Y,TY)=0$ one deduces that $w=w(x,z),
\bar{w}=\bar{w}(x,\bar{z})$. In other words, the Jacobian for the
coordinate change takes the following ``upper triangular''
form\footnote{Here we write $z$ to denote both $z$ and $\bar{z}$
coordinates to simplify the notation. The same applies to $w$.}
$$
\frac{\part (x,z,y)}{\part(u,w,v)} = \begin{pmatrix}{\part x}/{\part u} & {\part z}/{\part u}&{\part y}/{\part u}\\
0&{\part z}/{\part w}&{\part y}/{\part w}\\0&0&{\part y}/{\part v}\end{pmatrix}.
$$
It immediately follows that the $e_\mu$'s transform among themselves in a simple way:
$$e'_\mu = \frac{\part y^\nu}{\part v^\mu}e_\nu.$$
The transformation law for the $e_i$'s is slightly more
complicated. In the new chart, we have
$$e'_i = \frac{\part}{\part w^i}-i\om'_{ij}dw^j-i\om'_{ia}du^a.$$
The form of the Jacobian implies that the components of $\om$
transform according to
\begin{eqnarray}
\om'_{ij} &=& \frac{\part z^k}{\part w^i}\frac{\part z^\ell}{\part w^j}\,\om_{k\ell}\no\\
\om'_{ia} &=& \frac{\part z^j}{\part w^i}\frac{\part x^b}{\part u^a}\,\om_{jb}
+ \frac{\part z^j}{\part w^i}\frac{\part z^k}{\part u^a}\,\om_{jk}
+  \frac{\part y^\mu}{\part w^i}\frac{\part x^b}{\part u^a}\,\om_{\mu b}\no
\end{eqnarray}
Combining this with the transformation law for the coordinate
basis, one can show that
$$e'_i = \frac{\part z^j}{\part w^i}\, e_j + \frac{\part y^\mu}{\part w^i}\, e_\mu.$$
These is the same gluing cocycle should we take
\begin{equation}\label{obvtriv}
\tilde{e}_i=\frac{\part}{\part z^i}, \qquad \tilde{e}_\mu =
\frac{\part}{\part y^\mu}
\end{equation}
as the obvious local
trivialization of $\cL Y_\CC\oplus \cF Y^{1,0}$. Therefore our choice
of basis establishes an isomorphism between $E_Y$ and $\cL Y_\CC\oplus
\cF Y^{1,0}$  as vector bundles.

One can further show that these basis sections of $E_Y$ commute
under the Lie bracket derived from the Dorfman bracket on
$TX\oplus TX^\vee$. For instance, we have
\begin{eqnarray}
[e_i,e_j] &=& -i \cL_{\part_i}(\om_{j\al}dq^\al) +i \cL_{\part_j}(\om_{i\al}dq^\al)-d\iota_{\part_j}(i\om_{i\al}dq^\al)\no\\
&=& -i(\part_i\om_{j\al}-\part_j\om_{i\al}+\part_\al\om_{ij})dq^\al\no\\
&=& 0,
\end{eqnarray}
where $q^\al$ denote all of $x^a, z^i, \bar{z}^\bi, y^\mu$. The
last step follows directly from $d\om=0$. By the same token, we
have $[e_\mu,e_\nu]=0$, $[e_i,e_\mu]=0$. Namely, our basis
sections for $E_Y$ have the same (vanishing) Lie brackets among
themselves, just as the basis sections Eq.~(\ref{obvtriv})  of
$\cL Y_\CC\oplus \cF Y^{1,0}$.

This shows that $E_Y\simeq \cL Y_\CC\oplus \cF Y^{1,0}$ not only as
vector bundles but also as Lie algebroids. Since Lie algebroid
structures on $E\to X$ are in one-to-one correspondence with
degree-one homological vector fields on $E[1]$ (i.e. BRST operators in the
jargon of TFT), we conclude that one can use the obvious
Lie algebroid structure on $\cL Y_\CC\oplus \cF Y^{1,0}$ to compute
the open string ground states for a coisotropic A-brane. Namely one can
use the simplified BRST operator\footnote{We have changed $\part$
to $\bar\part$ to conform with the usual conventions.}
$$Q_Y = d_{\cL Y} + \bar{\part}_{\cF Y}$$
where $d_{\cL Y}$ is the de Rham differential in the leaf
direction, and $\bar{\part}_{\cF Y}$ is the Dolbeault operator in
the directions transverse to the foliation. This proves the claim
by D. Orlov and one of the authors~\cite{KO} that the open-string
BRST cohomology for a coisotropic A-brane is isomorphic to the
cohomology of the sheaf of functions locally constant along the
leaves of the characteristic foliation of $Y$ and holomorphic in
the transverse directions.

\section*{Acknowledgments}
A. K. would like to thank Alexey Bondal, Andrei C\u ald\u araru, Tony
Pantev, and Marco Gualtieri for helpful discussions. Y.L. would like to
thank Yong-Geun Oh for an interesting discussion. We are also
grateful to the organizers of the Workshop on Mirror Symmetry 
at the Perimeter Institute, Waterloo, for providing a
stimulating atmosphere. This work was supported in part by the DOE
grant DE-FG03-92-ER40701.

\end{document}